\documentclass[english,aps,amsmath,amssymb,prl,twocolumn]{revtex4}
\usepackage[T1]{fontenc}
\usepackage[utf8]{inputenc}
\usepackage{float}
\usepackage{amsmath}
\usepackage{graphicx}

\makeatletter

\providecommand{\tabularnewline}{\\}

\@ifundefined{textcolor}{}
{%
 \definecolor{BLACK}{gray}{0}
 \definecolor{WHITE}{gray}{1}
 \definecolor{RED}{rgb}{1,0,0}
 \definecolor{GREEN}{rgb}{0,1,0}
 \definecolor{BLUE}{rgb}{0,0,1}
 \definecolor{CYAN}{cmyk}{1,0,0,0}
 \definecolor{MAGENTA}{cmyk}{0,1,0,0}
 \definecolor{YELLOW}{cmyk}{0,0,1,0}
 }

\usepackage{amsfonts}
\makeatother

\usepackage{babel}
\begin{document}

\title{The Difficulty of Gate Control in Molecular Transistors }

\author{D. Hou}

\address{Department of Physics, Renmin University of China, Beijing 100872,
P. R. China}

\address{Department of Physics, Shandong University, Jinan 250100, P. R. China}

\author{J. H. Wei}

\email{wjh@ruc.edu.cn}

\address{Department of Physics, Renmin University of China, Beijing 100872,
P. R. China}
\begin{abstract}
The electrostatic gating effects on molecular transistors are investigated
using the density functional theory (DFT) combined with the nonequilibrium
Green's function (NEGF) method. When molecular energy levels are away
from the Fermi energy they can be linearly shifted by the gate voltage,
which is consistent with recent experimental observations {[}Nature
\textbf{462}, 1039 (2009){]}. However, when they move near to the
Fermi energy (turn-on process), the shifts become extremely small
and almost independent of the gate voltage. The fact that the conductance
may be beyond the gate control in the {}``ON'' state will challenge
the implementation of molecular transistors.
\end{abstract}


\maketitle
\emph{Introduction}. --- The molecular transistor is the most essential
element of molecular electronics. How to control molecular orbitals
by external gate voltage ($V_{\mathrm{G}}$) is the key problem of
molecular transistors. Although two-terminal molecular devices (with
the source and drain electrodes) have been extensively studied for
many years, the real three-terminal ones (with the gate electrode
added in) have only been developed recently in the laboratory by Song
\textit{et al} \cite{Son09}. They found that the transport current
of Au/1,4-benzenedithiol (BDT)/Au (as well as Au/1,8-octanedithiol/Au)
in a transistor structure can be directly controlled by gate modulating
molecular orbitals. A linear relationship between the gate voltage
and the shift of the molecular orbital energy has been also observed
\cite{Son09}. Based on their experiment, it seems that the practical
molecular field-effect-transistor (FET) will be realized soon.

From a practical point of view, in order to control the carriers or
the \textquotedbl{}conductive channel\textquotedbl{} by the gate voltage,
the molecular FET should work well on three modes of operation (as
the traditional counterpart does): cutoff, linear and saturation regions.
We comment that the gate control of molecular orbital only observed
in the cutoff region, i.e. the channel between the source and drain
keeping in the {}``OFF'' state in the experiment \cite{Son09}.
Of course, one expects molecular-orbital-gating occurring in the cutoff
region is also valid in the other two regions (the {}``ON'' states).
However, such natural extrapolation in micro electrons is questionable
in molecular transistors, because at the molecular level the motion
of carriers obeys quantum mechanical which may make the real situation
completely different. In this letter, we will address this issue via
the first principles calculations--the density functional theory (DFT)
combined with the nonequilibrium Green's function (NEGF) method. We
will demonstrate that above extrapolation does be problematic: once
the channel between the source and drain being turned on, the conductance
of the molecular transistor may be beyond the control of gate voltage.
This may be a serious challenge to the implementation of molecular
transistors. 

\emph{Method}. --- The gate-controlled Au/BDT/Au structure is the
object of our theoretical study. However, the aim of the present work
is not to quantitatively simulate the experment of Song \textit{et
al} , but to qualitatively reveal the physics behind it and the physics
being missed in it. Based on this point, we apply the gate voltage
via an additional external potential instead of adding a real gate
electrode in the first principle calculations (the latter is very
difficult for DFT+NEGF calculations).

\begin{figure}[!h]
\includegraphics[width=8cm]{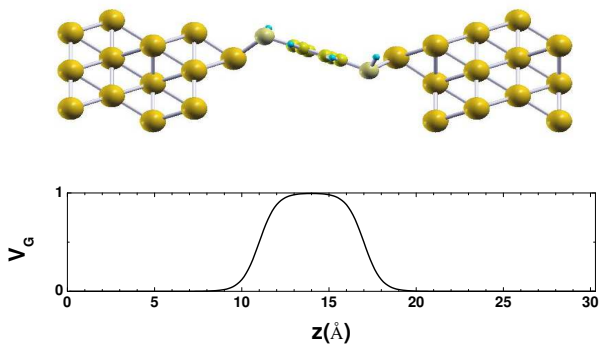} \caption{(color online) Upper panel: structure geometry of Au/1,4-benzenedithiol
(BDT)/Au junctions (the scattering region). Lower panel: $z$ dependent
gate voltage profile in the form shown in Eq.\eqref{eq:uz}.}

\label{figbdt} 
\end{figure}

The Au/BDT/Au structure consists three parts: the left and right electrodes
plus the scattering region (with gate voltage being applied). Both
of the electrodes are periodically repeated structures of the super
cell formed by 3 layers of 3$\times$3 Au(111) slabs, while the scattering
region is composed of a single molecule BDT and four layers of 3$\times$3
Au(111) slabs on each side. The BDT molecule bounds to the electrode
surfaces through the thiol groups contacting with Au adatoms on each
side, as shown in Figure \ref{figbdt}. It is reported that when BDT
molecule contacts with Au leads, the structure with the thiol group
S-H non-dissociated is energetically more\textbf{ }favorite\cite{Ji09,Strange11},
thus we adopted this kind of structure in our calculation. As in most
literatures, BDT molecule is set perpendicular to the Au surfaces.
Although the molecule structure and its coupling to the electrodes
may change the conductance of BDT single molecular junction, they
should not cause qualitatively difference to the results reported
here. We will\textbf{ }neglect those effects and focus on the gate-controlled
conductance of single-structure molecular junctions.

All of the calculations are performed with spin-polarized DFT implemented
in the Siesta code \cite{siesta1,siesta2} using Perdew-Burke-Ernzerhof
exchange-correlation functional \cite{PBE}. Double-$\zeta$ plus
polarization function(DZP) basis set is used, together with a mesh
cutoff of 200 Ry and norm-conserving pseudopotentials. The distance
between the left and right lead surfaces is relaxed to the equilibrium
position. Then the BDT molecule together with the nearest layer of
Au atoms are fully relaxed to fulfill the energy and force convergences
of 10$^{-5}$ eV and 0.01 eV/Å respectively. 

The transport property is calculated using DFT+NEGF method implemented
in the recent Siesta code (the TranSiesta part \cite{Bra02}). The
periodic boundary conditions are used in directions orthogonal to
the transport direction. A 3$\times$3 uniform $k$ meshes are sampled
over this two-dimensional Brillouin zone. The density is integrated
with 32 points on the semicircle, 32 points parallel to the real axis,
and 16 poles. The transmission spectrum at zero bias is obtained by
integrating over 1000 points. 

To study the electrostatic gating effects, we follow the idea of Morari
\emph{et al.}, i.e. the gate voltage is simulated by adding external
potential determined solely by the coordinate $z$ along the transport
direction \cite{morari09}. In our work, the external potential is
in the form of 
\begin{equation}
U(z)=eV_{\mathrm{G}}(\frac{1}{1+e^{(z-z_{1})/\Delta z}}-\frac{1}{1+e^{(z-z_{2})/\Delta z}})\label{eq:uz}
\end{equation}
where $\Delta z=0.5$ Å. The parameters $z_{1}$ and $z_{2}$ are
used to ensure that the applied gate-potential lies in the molecule
region and decrease to zero smoothly when reaching the electrodes,
as the potential profile shown in Figure \ref{figbdt}. Comparing
with experiments, this simulation of gate voltage obviously lacks
the effect of the gate dielectric ($\mathrm{Al_{2}}\mathrm{O}_{3}$
layer in Ref.\onlinecite{Son09}). This effect should not distinctly
change the physics presented in this work, at least in the qualitatively
level. 

\begin{figure}[!h]
\includegraphics[width=8cm]{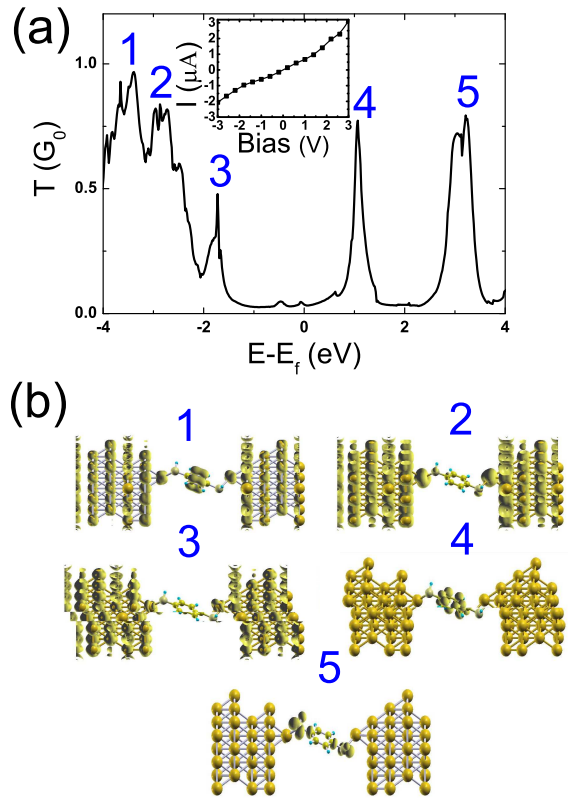}

\caption{(color online) (a) The transmission spectrum $T(E)$ for BDT junction
under zero gate voltage in the unit of quantum conductance $G_{0}$($\equiv2e^{2}/h$)
. The insert is the calculated $I-V$ curve. (b) The spatial charge
density within energy ranges around four transmission peaks labeled
by 1-5 in (a).}

\label{fig0vg} 
\end{figure}

\emph{Results and discussions} . --- We first check our calculation
before the gate voltage being applied and the results\textbf{ }are
summarized in Figure \ref{fig0vg}, where the upper panel depicts
the zero-bias transmission spectrum $T(E)$ (the Fermi energy $E_{\mathrm{F}}$
is set as the zero point of energy), with its insert showing the calculated
$I-V$ curve. One can see that the current is of the same order ($\mathrm{\mu}\mathrm{A}$)
with experimental measurement \cite{Son09}, which is apparently an
acceptable result for static DFT. It further confirms the chosen junction
geometry (see Figure \ref{figbdt}) is reasonable. The asymmetric
character of $I-V$ curve is also consistent with experiment \cite{Son09},
which may result from the asymmetric coupling of BDT molecule with
the left and right electrodes. 

\begin{figure}[!h]
\includegraphics[width=8cm]{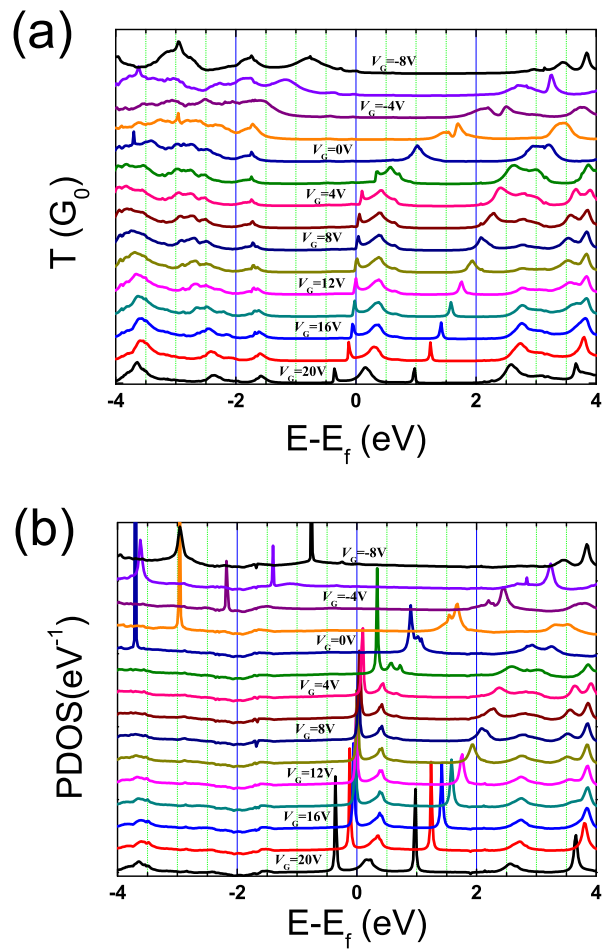} \caption{(color online) Calculated transmission spectrum $T(E)$ (a) and projected
density of states (PODS) on carbon atoms (b) for BDT junction under
various gate voltages. }

\label{fig-gz} 
\end{figure}

From Figure \ref{fig0vg}, we find five transmission peaks within
the energy range of $-4$ to $4$ eV and they are labeled as peaks
1-5, where peaks 1-3 (centered on $-3.7$, $-2.9$ and $-1.9$ eV
respectively) correspond to hole-transport while peaks 4-5 (centered
on $1.2$ and $3.0$ eV respectively) to electron-transport. Actually,
we can further identify which element (C, S or Au) contributes most
to those peaks via the spatial charge density {[}also called the local
density of states (LDOS){]} within energy ranges around them, as shown
in the lower panel of Figure \ref{fig0vg}. It can be seen that C
atoms contribute most to both peaks 1 and 4, while Au atoms (including
those linking BDT and bulk electrodes) do most to peaks 2 and 3. The
main contribution to peak 5 is clearly from S atoms and the S-C hybridization.
From those information, we can conclude that peaks 1 and 4 are respectively
result from the conduction of the highest occupied molecular orbital
(HOMO) and lowest unoccupied molecular orbital (LUMO) of BDT. Since
the LUMO is nearest to $E_{\mathrm{F}}$, it will dominate the transport
properties of the device under low bias voltage. This result seems
not consistent with the experiment of Song \emph{et al}, where the
HOMO is observed to be turned on prior to the LUMO. It indicates that
some factors may be missed in our ideal DFT calculations, e.g. many
body corrections \cite{Strange11}. We argue that the physics of activating
and controlling LUMO or HOMO should be same for molecular transistors
\footnote{Actually, that point has been proved by Song \emph{et al }in their
experiment, e.g. see the Supplementary Fig. 8 of Ref.\onlinecite{Son09}.
In the present paper, the evidence is shown in Figure \ref{fig-gz}(b). %
}. 

Now, let us show the complete story when gate voltage being applied
in the form of Eq.\eqref{eq:uz} (cf. Figure \ref{figbdt}). In Figure
\ref{fig-gz}(a), we plot the transmission spectrum under various
gate voltages (from$-8$ V to $+20$ V with the interval of $2$ V).
As we have demonstrated in Figure \ref{fig0vg}, the LUMO of BDT plays
the leading role in the molecular-orbital-gating process under low
bias. It should be enough for us to focus on the gate control of this
orbital, e.g. the change of transmission peak 4 with $V_{\mathrm{G}}$.
To be more specific, Figure \ref{fig-gz}(b) depicts the projected
density of states (PDOS) only on carbon atoms (to highlight the change
of LUMO) under synchronously changing $V_{\mathrm{G}}$ as shown in
Figure \ref{fig-gz}(a). Combining Figure \ref{fig-gz}(a) and \ref{fig-gz}(b),
we can clearly see the modulation of $V_{\mathrm{G}}$ on the energy
of the LUMO ($E_{\mathrm{L}}$) and accurately understand its mechanism. 

At first glance, Figure \ref{fig-gz} does confirm the fact that the
conductance and molecular orbital of BDT junction can be directly
controlled by $V_{\mathrm{G}}$. However, in different range of $V_{\mathrm{G}}$,
molecular-orbital-gating exhibits different behavior patterns. Generally
speaking, Figure \ref{fig-gz} {[}both (a) and (b){]} can be divided
into three regions in term of $V_{\mathrm{G}}$, which are explained
in details as follows.

\begin{figure}[!h]
\includegraphics[width=8cm]{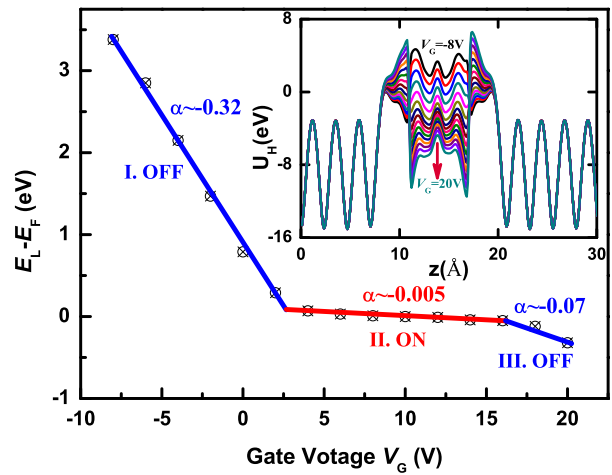} \caption{(color online) The change of molecular orbital (LUMO of BDT) energy
as a function of the gate voltage. The straight lines are eye-guide
lines with different slopes.  The insert is the $x-y$ planar-averaged
converged Hartree potentials under gate voltage increasing from$-8$
V to $+20$ V with the interval of $2$ V, as indicated by the arrow.
The difficulty of gate control in molecular transistors in the {}``ON''
state is clearly shown in the figure. }

\label{figde} 
\end{figure}

Region I ($-8\:\mathrm{V}<$$V_{\mathrm{G}}<3\mathrm{\: V}$): In
this region, the molecular transistor is in the {}``OFF'' state
due to the near zero DOS and near zero $T(E)$ at $E_{\mathrm{F}}$.
Actually, it is the region where Song\emph{ et al} observed molecular-orbital-gating
in their experiment \cite{Son09}. Our theoretical results verify
their observation that the gate control of molecular orbital is linear
and effective. To demonstrate this point in more details, we extract
the change of the energy of the LUMO (defined as $E_{\mathrm{L}}-E_{\mathrm{F}}$)
as functions of $V_{\mathrm{G}}$ from Figure \ref{fig-gz} (in all
range of $V_{\mathrm{G}}$, not only in Region I), and summarize the
results in Figure \ref{figde}. In that figure, the remarkable linear-modulation
of $V_{\mathrm{G}}$ on $E_{\mathrm{L}}$ is clearly shown in region
I, with the slope being defined as the gate efficiency factor, $\alpha\equiv\Delta E_{\mathrm{L}}/e\Delta V_{\mathrm{G}}$
(cf. Ref.\onlinecite{Son09}). From Figure \ref{figde}, we work out
$\alpha\sim-0.32$ in Region I.

By referring \ref{fig-gz}(b), one can find a similar linear-control
of $V_{\mathrm{G}}$ on the energy of the HOMO ($E_{\mathrm{H}}$)
in the energy range of $E<E_{\mathrm{F}}$. The gate efficiency factor,
$\alpha_{\mathrm{H}}\;(\equiv\Delta E_{\mathrm{H}}/e\Delta V_{\mathrm{G}})\sim0.32$,
is the same as that of gate control of $E_{\mathrm{L}}$ in spite
of a sign difference. That result confirms our argument that activating
and controlling LUMO or HOMO should share the same physics. Not surprisingly,
the activation of the HOMO is much harder than that of the LUMO in
our theory. 

Our theoretical $\left|\alpha\right|\;(\sim0.32$) is comparable with
the experimental value $\alpha\sim0.22$ \cite{Son09}. That provides
a reliable basis for further understanding theoretical results in
Region II, which has not been reported in experiments yet. 

Region II ($3\:\mathrm{V}<$$V_{\mathrm{G}}<16\mathrm{\: V}$): In
this region, the molecular transistor is in the {}``ON'' state due
to the finite DOS and finite $T(E)$ at $E_{\mathrm{F}}$. As shown
in Figure \ref{fig-gz} and \ref{figde}, the behavior of molecular-orbital-gating
in this region is completely different from that in Region I: both
DOS and $T(E)$ near $E_{\mathrm{F}}$ change very slowly with further
increasing $V_{\mathrm{G}}$ once the LUMO has been activated to be
conductive. If we still linearly fit the gate modulating on $E_{\mathrm{L}}$
in Figure \ref{figde}, we will get a near zero $\alpha$ ($\sim-0.005$).
It indicates that the energy of LUMO becomes beyond the control of
gate voltage in this region. 

The weak dependence of $T(E)$ on $V_{\mathrm{G}}$ means the source-drain
current $I_{\mathrm{SD}}$ at finite bias also independent on $V_{\mathrm{G}}$,
i.e. $V_{\mathrm{G}}$ will lose the control of $I_{\mathrm{SD}}$
in molecular transistors in the {}``ON'' state. That will challenges
the implementation of molecular transistors. As we know, $I_{\mathrm{SD}}$
at certain source-drain bias ($V_{\mathrm{SD}}$) is supposed to increase
with $V_{\mathrm{G}}$ linearly in the linear region and quadratically
in the saturation region. Neither of them can be achieved when the
molecular orbital being beyond the control of $V_{\mathrm{G}}$. 

In order to gain an insight into the difficulty of gate control in
Region II, we show the $x-y$ planar-averaged converged Hartree potentials
($U_{\mathrm{H}}$) under various $V_{\mathrm{G}}$ in the insert
of Figure \ref{figde}, where the peaks correspond to chemical bonds
while the valleys to chemical groups (CH or SH) or Au atoms. By comparing
Figure \ref{figde} and Figure \ref{figbdt}, we find two potential
nodes at the interface between BDT and Au electrodes (S-Au bonds).
The potential of S atoms inside the nodes drops rapidly with increasing
$V_{\mathrm{G}}$, while that of Au atoms outside of the nodes grows
fast. 

Since negative potential tends to trap electrons but positive one
trap holes, those two nodes indicate that $V_{\mathrm{G}}$-induced
negative and positive charges localized at the interfaces separates
from each other (see Table I for further details). As a consequence,
an inner electric field builds up to offset the $V_{\mathrm{G}}$-induced
one. This process is very similar to the shielding of external field
by a plate capacitance, we thus called the effective capacitance near
the (left or right) interface as structural capacitance ($C_{S}$).
It can explain why the effective potential acting on molecular energy
is less than the applied one ($\left|\alpha\right|<1$), even no gate
dielectric structure being included. 

\begin{table}[H]
{\small \caption{Gate voltage $V_{\mathrm{G}}$ induced changes of charges at left
($\mathrm{\mathrm{Au}_{L}-S}_{L}$) and right interfaces ($\mathrm{S}_{R}-\mathrm{Au}_{R}$)
and that within the total scattering region. $V_{\mathrm{G}}$ varies
from$-8$ V to $20$ V with the interval of $4$ V.}
}{\small \par}

{\small \label{tab1}\centering }{\small \par}

{\small }%
\begin{tabular}{c|cccccccc}
\hline 
$V_{\mathrm{G}}$ & -8 & -4 & 0 & 4 & 8 & 12 & 16 & 20\tabularnewline
\hline 
$\mathrm{S}_{L}$ & 0.09 & 0.04 & 0 & -0.03 & -0.07 & -0.11 & -0.16 & -0.21\tabularnewline
\hline 
$\mathrm{Au}_{L}$ & -0.13 & -0.06 & 0 & 0.08 & 0.18 & 0.27 & 0.37 & 0.46\tabularnewline
\hline 
$\mathrm{S}_{R}$ & 0.11 & 0.05 & 0 & -0.06 & -0.11 & -0.17 & -0.23 & -0.29\tabularnewline
\hline 
$\mathrm{Au}_{R}$ & -0.16 & -0.07 & 0 & 0.08 & 0.18 & 0.29 & 0.40 & 0.50\tabularnewline
\hline 
{\small Total} & 0.67 & 0.32 & 0 & -0.41 & -0.94 & -1.48 & -2.02 & -2.54\tabularnewline
\hline 
\end{tabular}
\end{table}

When the molecular transistor is in the {}``ON'' state, the finite
DOS at $E_{\mathrm{F}}$ {[}$D(E_{\mathrm{F}})${]} will lead to the
so called quantum capacitance effects \cite{QC,Datta}. The quantum
capacitance $C_{Q}$ can be defined through $C_{Q}=e^{2}D(E_{\mathrm{F}})$.
In a realistic transistor, the effective capacitance $C$ is the series
combination of the electrostatic capacitance ($C_{E}$) and $C_{Q}$.
In our theory, $C_{E}$ is replaced by structural capacitance $C_{S}$,
therefore, $C=C_{S}C_{Q}/(C_{S}+C_{Q})$, which is obviously dominated
by the smaller of the two. As shown in the insert of Figure \ref{figde},
the narrow distance between opposite charges indicates that $C_{S}$
should be very large (plate capacitance model), which makes $C_{Q}$
dominate. As a consequence, $C_{Q}$ effectively reduces the external
potential acting on BDT, or even completely shield $V_{\mathrm{G}}$-induced
field to make $V_{\mathrm{G}}$ loss control of molecular orbitals. 

Region III ($16\:\mathrm{V}<$$V_{\mathrm{G}}<20\mathrm{\: V}$):
The molecular transistor is turned back to {}``OFF'' state when
sufficient large $V_{\mathrm{G}}$ shifts the LUMO of BDT away from
$E_{\mathrm{F}}$. The efficient linear-modulation of $V_{\mathrm{G}}$
on molecular orbitals is also restored but with a different slope
($\alpha\sim-0.07$) from that in Region I. When the LUMO+1 of BDT
is shifted near $E_{\mathrm{F}}$, one can expect a similar process
as happens in Region II. In that case, the electron-electron correlation
should be carefully treated (beyond the mean-field theory) due to
more electrons being trapped within the BDT molecule, which has been
already beyond the scope of the present work.

\emph{Summary}. --- In summary, we have studied the gate control of
molecular orbitals by DFT+NEGF method. When molecular energy levels
are away from the Fermi energy they can be linearly shifted by the
gate voltage, which is consistent with recent experimental observations.
When they move near to the Fermi energy, the shifts become extremely
small and almost independent of the gate voltage, which indicates
that it may be difficult for the gate voltage to control molecular
orbitals in the {}``ON'' state.
\begin{acknowledgments}
Support from NSFC of China (Grant No. 11074303) are gratefully acknowledged.
Part of the computer time is supported by Physics Laboratory for High
Performance Computing, Renmin University of China. \end{acknowledgments}

\end{document}